\documentclass[10pt,conference]{IEEEtran}
\IEEEoverridecommandlockouts

\usepackage{cite}

\usepackage{amsmath,amssymb,amsfonts}
\usepackage{algorithmic}
\usepackage{graphicx}
\usepackage{textcomp}
\usepackage{xcolor}
\def\BibTeX{{\rm B\kern-.05em{\sc i\kern-.025em b}\kern-.08em
    T\kern-.1667em\lower.7ex\hbox{E}\kern-.125emX}}
\usepackage{tcolorbox}
\usepackage{bbding}
\usepackage{multirow}
\usepackage{multicol}
\usepackage{booktabs}
\usepackage{array}
\usepackage{colortbl}  
\usepackage{xcolor}
\usepackage{arydshln}
\usepackage{xspace}
\usepackage{hyperref}

\newcommand{\authornote}[3]{{\color{#2} #3}}

\newcommand{\ylrevised}[1]{#1}
\newcommand\dewu[1]{\authornote{dewu}{black}{#1}}
\newcommand{\later}[1]{}

\newcommand{\HumanEvo}{HumanEvo\xspace}

\newcommand{\boxmargin}{1mm}
\tcbuselibrary{most, skins, breakable, theorems}
\newtcolorbox{myboxa}[2][]{
    colback=gray!10!white,
    colframe=black, enhanced,
    attach boxed title to top left={yshift=-2mm,xshift=5mm},
    title=#2,#1
}
\newtcolorbox{myboxb}[2][]{
    boxsep=1pt,
    left = \boxmargin, right = \boxmargin, top = \boxmargin, bottom = \boxmargin,
    title={#2},#1
}
\newtcolorbox{myboxc}{
    colback=gray!15!white,
    arc = 0pt, outer arc = 0pt,
    boxsep=0pt, left = 3pt, right = 0pt, top = 0pt, bottom = 0pt, 
    leftrule=3pt, bottomrule=0pt,toprule=0pt, rightrule=0pt,
    left = \boxmargin, right = \boxmargin, top = \boxmargin, bottom = \boxmargin
}

\begin{document}

\title{HumanEvo: An Evolution-aware Benchmark for More Realistic Evaluation of Repository-level Code Generation}



\author{
    \IEEEauthorblockN{Dewu Zheng$^{1}$, Yanlin Wang$^{1*}$\thanks{* Yanlin Wang is the corresponding author.}, Ensheng Shi$^{2}$, Ruikai Zhang$^{3}$, Yuchi Ma$^{3}$, Hongyu Zhang$^{4}$, Zibin Zheng$^{1,5}$}
    \IEEEauthorblockA{$^{1}$ Sun Yat-sen University, Zhuhai, China}
    \IEEEauthorblockA{
    \href{mailto：zhengdw5@mail2.sysu.edu.cn}{zhengdw5}@mail2.sysu.edu.cn,
    \{\href{mailto:wangylin36@mail.sysu.edu.cn}{wangylin36}, 
    \href{mailto:zhzibin@mail.sysu.edu.cn}{zhzibin}\}@mail.sysu.edu.cn,
    }
    \IEEEauthorblockA{$^{2}$ Huawei Cloud Computing Technologies Co., Ltd., Beijng, China}
    \IEEEauthorblockA{
    \href{mailto:shiensheng@huawei.com}{shiensheng}@huawei.com,
    }
    \IEEEauthorblockA{$^{3}$ Huawei Cloud Computing Technologies Co., Ltd., Shenzhen, China}
    \IEEEauthorblockA{
    \{\href{mailto:zhangruikai1@huawei.com}{zhangruikai1}, \href{mailto:mayuchi1@huawei.com}{mayuchi1}\}@huawei.com
    }
    \IEEEauthorblockA{$^{4}$ Chongqing University}
    \IEEEauthorblockA{
    \href{mailto:hyzhang@cqu.edu.cn}{hyzhang}@cqu.edu.cn
    }
    \IEEEauthorblockA{$^{5}$ Zhuhai Key Laboratory of Trusted Large Language Models}
}

\maketitle

\begin{abstract}

To evaluate the repository-level code generation capabilities of Large Language Models (LLMs) in complex real-world software development scenarios, many evaluation methods have been developed. These methods typically leverage contextual code from the latest version of a project to assist LLMs in accurately generating the desired function. However, such evaluation methods fail to consider the dynamic evolution of software projects over time, which we refer to as evolution-ignored settings. This in turn results in inaccurate evaluation of LLMs' performance. In this paper, we conduct an empirical study to deeply understand LLMs' code generation performance within settings that reflect the evolution nature of software development. To achieve this, we first construct an evolution-aware repository-level code generation dataset, namely \HumanEvo, equipped with an automated execution-based evaluation tool. Second, we manually categorize \HumanEvo according to dependency levels to more comprehensively analyze the model's performance in generating functions with different dependency levels. Third, we conduct extensive experiments on \HumanEvo with seven representative and diverse LLMs to verify the effectiveness of the proposed benchmark. We obtain several important findings through our experimental study. For example, we find that previous evolution-ignored evaluation methods result in inflated performance of LLMs, with performance overestimations ranging from 10.0\% to 61.1\% under different context acquisition methods, compared to the evolution-aware evaluation approach. Based on the findings, we give actionable suggestions for more realistic evaluation of LLMs on code generation. We also build a shared evolution-aware code generation toolbox to facilitate future research. The replication package including source code and datasets is available at \url{https://github.com/DeepSoftwareAnalytics/HumanEvo}.

\end{abstract}

\section{Introduction}

In recent years, the LLM-based code generation task has drawn widespread attention~\cite{survey1,survey2,language,dong2023self,jiang2023selfevolve,zhang2023planning,metagpt,chatdev, wang2024beyond,wang2024sparsecoder}. Many code LLMs~\cite{codellama, starcoder, deepseek, CodeGeeX,palm,allal2023santacoder,wizardcoder, pangu,pangu2,codegen, codex, AlphaCode, incoder} are being employed to empower programming assistants, which are playing a vital role in practical software development~\cite{nie2023unveiling,gu2022accelerating,wei2023magicoder,wang2024RLCoder}. 
Recently, several repository-level code generation benchmarks such as CoderEval~\cite{codereval}, RepoBench~\cite{repobench}, EvoCodeBench~\cite{evocodebench}, and RepoEval~\cite{repocoder} have emerged to better simulate real-world development scenarios. They sample functions from real-world projects as programming tasks, aiming to reflect the performance of LLMs in actual development by prompting them to generate these functions according to the target function description and project context. 

Despite the fact that real-world projects are inherently dynamic and evolve over time~\cite{10.1145/1056018.1056025,1347427,1572323}, current code generation benchmarks often overlook this critical aspect. They tend to treat the latest version of the repository as context source for evaluation, a situation we term as the \emph{evolution-ignored} phenomenon. This evaluation approach fails to accurately capture true nature of real software development scenarios where projects continuously evolve over time, which means that the project context that developers face when writing different code changes constantly. 
Therefore, to accurately assess the repository-level coding capabilities of LLMs, it is imperative to factor in this evolution aspect. Specifically, when prompting LLMs to generate different target functions, the project context provided should mirror the one available to developers at the time of the target function's creation. 
Unfortunately, evolution-ignored evaluation in previous benchmarks provides LLMs with an inaccurate and potentially misleading programming environment, leading to serious issues. 
We have found concrete instances of issues that stem from the evolution-ignored setting in real GitHub projects, which are detailed in Section~\ref{motivating examples}.


\textbf{Benchmark \HumanEvo.} To fill this gap,    
we construct {\HumanEvo}, a novel evolution-aware repository-level code generation benchmark, which better simulates real-world development processes and reflects the evolution nature of projects over time. \dewu{The evaluation process of \HumanEvo is as follows: }to ensure that the context we provide to LLMs mirrors the context available to the programmer when writing the target function (neither including additional code introduced by subsequent version changes nor omitting code that may have been updated or deleted due to version change), we roll back the entire repository to the state before the target code was committed and evaluate the code generation performance of LLMs \dewu{at function granularity }on the rolled-back repository.


We conduct a rigorous data construction pipeline to construct \HumanEvo. We start with selecting high-quality projects and then collecting a large number of pull requests (PRs) from these projects. Then, we perform an initial filtering on the crawled PRs to meet three attributes to guarantee the quality of the newly added functions in PRs. \dewu{Additionally, to ensure the reliability of the \HumanEvo's test suite, we establish a runtime environment for each PR and execute the corresponding project's testing framework to verify that the newly added functions are covered by the test suite.} Finally, we obtain 400 task instances, 200 for Python and 200 for Java. For each task instance in \HumanEvo, we record its corresponding PR's metadata to roll back the repository to the state prior to the target function’s commit.

\textbf{Empirical Study.} Based on the new benchmark \HumanEvo, we perform the first study to reveal LLMs' repository-level code generation capability in the evolution-aware setting. In particular, we conduct an extensive evaluation of 7 mainstream LLMs, including both open-source ones (CodeLlama-7B, CodeLlama-13B, CodeLlama-34B~\cite{codellama}, DeepSeekCoder-6.7B, DeepSeekCoder-33B~\cite{deepseek}) and closed-source ones (GPT-4~\cite{gpt4} and GPT-3.5-Turbo~\cite{gpt3.5}) on \HumanEvo. 

According to the empirical results, we have the following findings:
\textcircled{1} We find that all the studied LLMs show much worse performance on our evolution-aware setting. Previous evolution-ignored evaluation would lead to the inflated performance of LLMs, ranging from 10.0\% to 61.1\%, revealing how previous evaluation methods provided the models with a deviated programming scenario from reality.
\textcircled{2} 
\ylrevised{We find that the impact of the evolution-ignoring setting is more severe for target code with more complex dependencies. Specifically, compared to the evolution-ignored setting, the success rate of LLMs in generating functions with intra-class dependencies under the evolution-aware setting has been observed to decrease by approximately 14.0\% on average, while for functions with inter-class dependencies, the decline is 30.9\% on average.} 
\textcircled{3} 
\ylrevised{We investigate how the performance of LLMs change as the repository evolve. Experimental results show that in the evolution-ignored setting, in general, models’ performance gradually deviates from the real performance as the project evolves.}
\textcircled{4} 
We find that\dewu{, under the evolution-ignored setting, utilizing docstrings in different styles as input consistently leads to inflated performance.} Experimental results indicate that LLMs' code generation performance improves when given a detailed docstring compared to a brief one.


This work makes the following key contributions:

\begin{itemize}
    \item We have identified a common flaw in previous repository-level code generation benchmarks: they are evolution-ignored, causing inaccurate simulations of real-world development scenarios.
    \item We introduce a new benchmark \HumanEvo, an evolution-aware repository-level code generation benchmark that addresses the evolution-ignored issue, to conduct a comprehensive empirical study on the performance of LLMs in repository-level code generation.
    \item Through extensive experimentation, we substantiate the evolution-ignored situation leads to inflated performance of the models, with LLMs showing a performance inflation ranging from 10.0\% to 61.1\%.

    \item For convenient usage of \HumanEvo, we release all data, code, and the evaluation platform with Docker images that provide runtime environments for all projects, facilitating automated execution of all test suites. The replication package is provided at \url{https://anonymous.4open.science/r/HumanEvo/}.
\end{itemize}


\begin{figure}[t]
  \centering
  \includegraphics[width=0.9\linewidth]{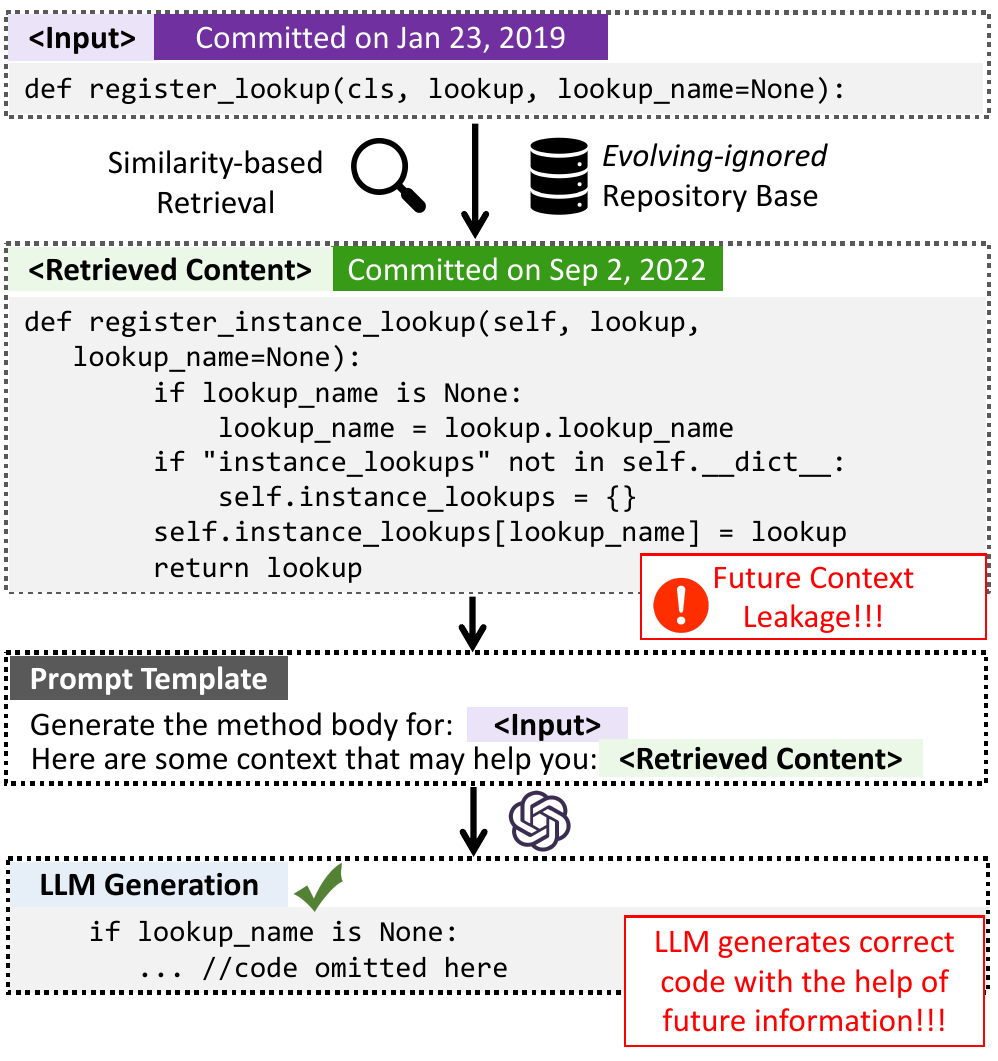}
  \caption{Example 1: future context leakage.}
  \label{fig:future_context_leakage}
\end{figure}

\begin{figure}[t]
  \centering
  \includegraphics[width=0.9\linewidth]{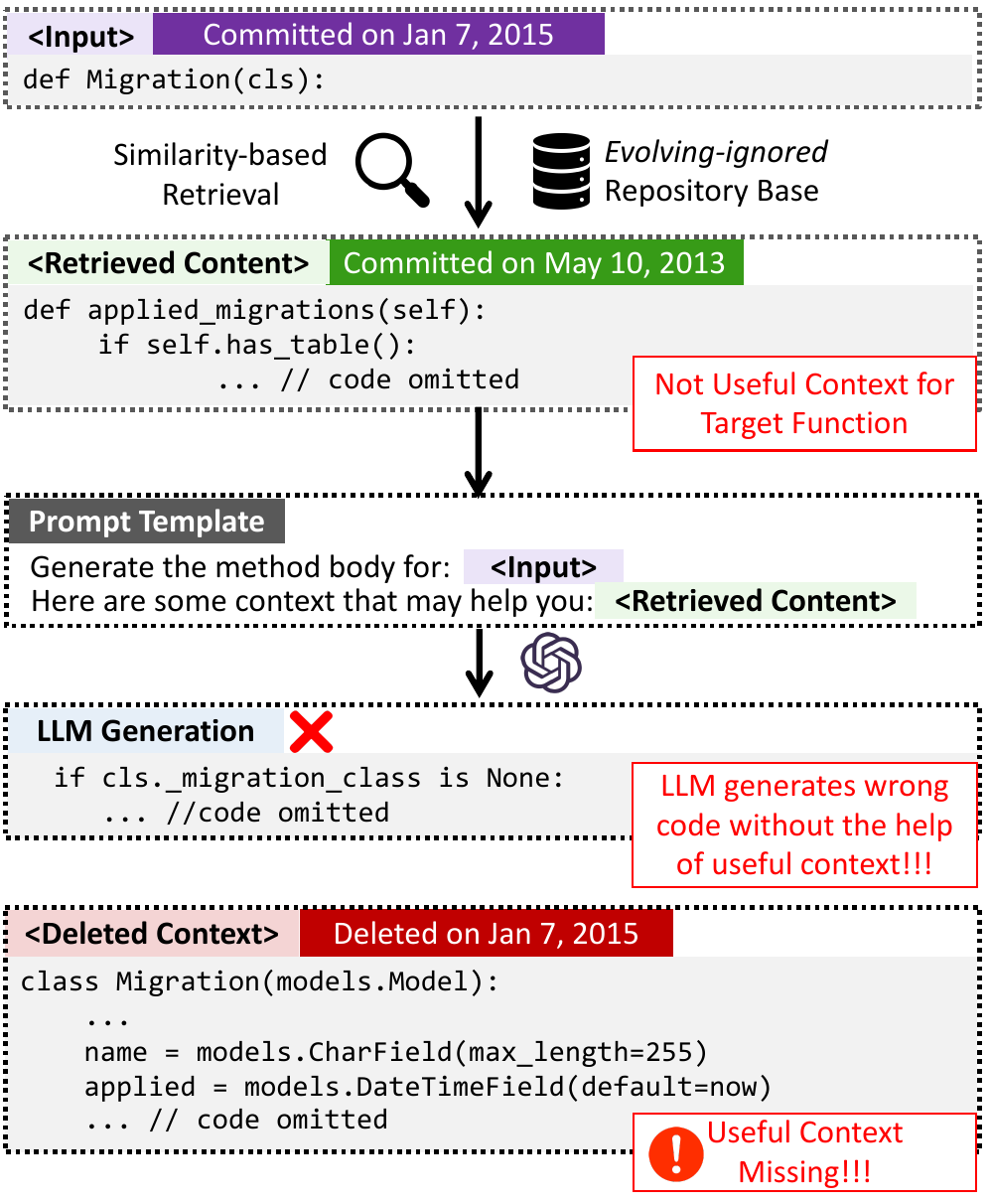}
  \caption{Example 2: useful context missing.}
  \label{fig:useful_context_missing}
\end{figure}

\section{Motivating Examples}
\label{motivating examples}

\ylrevised{
In this section, we firstly elaborate on how real repository evolves over time. Then, we present two motivating examples\footnote{The motivating examples are from the real-world project \url{https://github.com/django/django}} that showcase the specific issues that can emerge under the evolution-ignored setting.
}

\ylrevised{
In real-world development, projects undergo continuous evolution over time~\cite{negoita2019code,dabbish2012social,kalliamvakou2015open}. During the development phase of a project, programmers need to develop the project (i.e., by adding new features, removing existing code, adding test code, etc) based on the project requirements. In the maintenance phase, programmers continuously monitor the project and promptly fix any bugs that arise. These activities result in varying degrees of modifications to the project, causing the project to evolve over time~\cite{kallis2021predicting, rahman2014insight}. For a realistic and accurate evaluation of LLMs' code generation capabilities, this evolution nature should be considered. Otherwise, severe issues under the evolution-ignored setting would emerge. We have identified concrete instances of issues, which we refer to as \emph{future context leakage} and \emph{use context missing}, which we will illustrate with the following two examples.
}


\textbf{Example 1.}
\ylrevised{
Figure~\ref{fig:future_context_leakage} illustrates the \textbf{future context leakage} issue that can arise from the evolution-ignored setting. This example involves a target function that was originally committed on Jan 23, 2019. 
In the canonical RAG code generation pipeline, a retriever finds similar code from the corresponding repository (which is the latest version in the evolution-ignored setting) to help LLMs generate the target code. However, the retriever yields a function committed on Sep 2, 2022, a date that is \emph{later than the target function}. This process is illogical because it is implausible for a developer to access code that was not yet written at the time of the creation of the target function. Therefore, providing this code snippet as context to the model is a future context leak. While in this example the LLM generates the correct code, the reliability is questionable as it is uncertain whether it can still generate correctly without the help of this future code.  
}


\textbf{Example 2.}
\ylrevised{
Figure~\ref{fig:useful_context_missing} illustrates the \textbf{useful context missing} issue that may arise in the evolution-ignored setting. In this example, the function committed on Jan 7, 2015 is the target function. In the same RAG code generation pipeline as in Example 1, the retrieved code snippet is not particularly useful and LLM fails to generate the correct code. However, when the developer wrote that code, there was actually a piece of code in the repository that was highly similar to the target code and could have been retrieved as context, which would have likely helped generate the correct code. 
But this code snippet was deleted during the evolution process. As a result, under the evolution-ignored setting, this piece of code did not exist in repository context. While in this example the LLM generates wrong code, it is possible that providing LLM with this useful context might have enhanced its performance. 
}

\section{\HumanEvo Benchmark}
\label{sec:benchmark}
\ylrevised{
In this section, we introduce our new benchmark \HumanEvo in detail. We present the benchmark overview (Section~\ref{sec:benchmark_overview}), benchmark construction pipeline (Section~\ref{sec:benchmark_construction}), and benchmark characteristics (Section~\ref{sec:benchmark_characteristics}).
}

\subsection{Benchmark Overview}
\label{sec:benchmark_overview}

\begin{figure}[t]
  \centering
  \includegraphics[width=\linewidth]{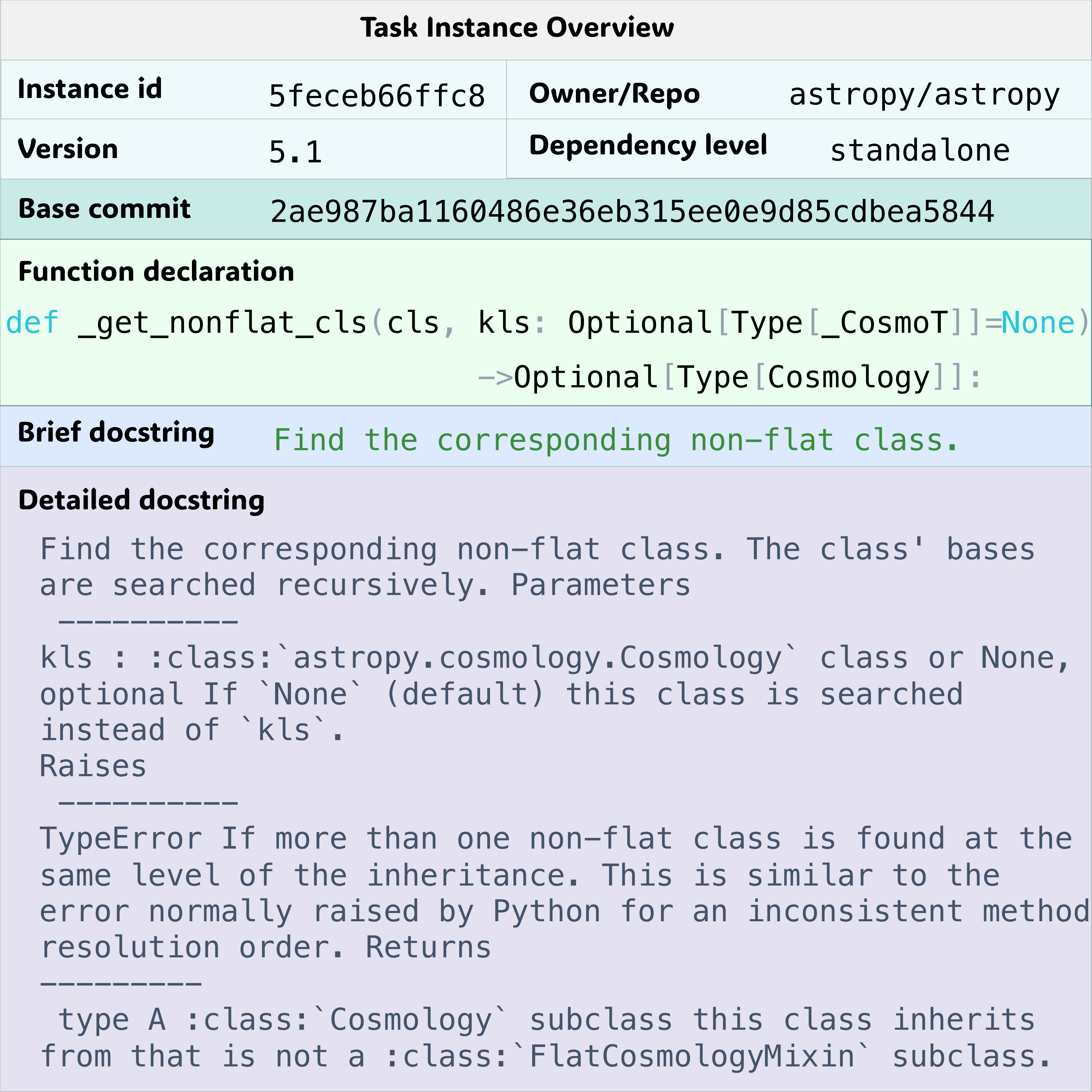}
  \caption{Task instance overview.}
  \label{fig:task_instance}
\end{figure}

\ylrevised{
\HumanEvo consists of a total of 400 task instances, comprising 200 Python programming tasks and 200 Java programming tasks, respectively. Figure~\ref{fig:task_instance} shows an example of a task instance overview. Each task instance is associated with a unique identification number (\texttt{Instance id}). The \texttt{Owner/Repo} field helps identify the GitHub repository from which the selected function originates, while the \texttt{Version} field indicates the version of the repository according to the implementation time of this function. \texttt{Base commit} refers to the last commit on the branch before the target function was committed. We use it to restore the repository to the state before the target function was implemented. The \texttt{Function declaration} of the target function will be provided to the LLM as part of the prompt, along with its corresponding docstring. Since the docstring styles of functions in pragmatic projects vary, with some adopting line-by-line comments and others even lacking corresponding comments, we think it is necessary to rewrite the docstring for each selected function. As depicted in Figure~\ref{fig:task_instance}, we provide two styles of docstrings. The \texttt{Brief docstring} provides a concise summary of the functionality of the target function, while the \texttt{Detailed docstring} provides a thorough description of the function's functionality, inputs, and outputs.
}

\subsection{Benchmark Construction Pipeline}
\label{sec:benchmark_construction}

\begin{figure*}[t]
  \centering
  \includegraphics[width=1.0\textwidth]{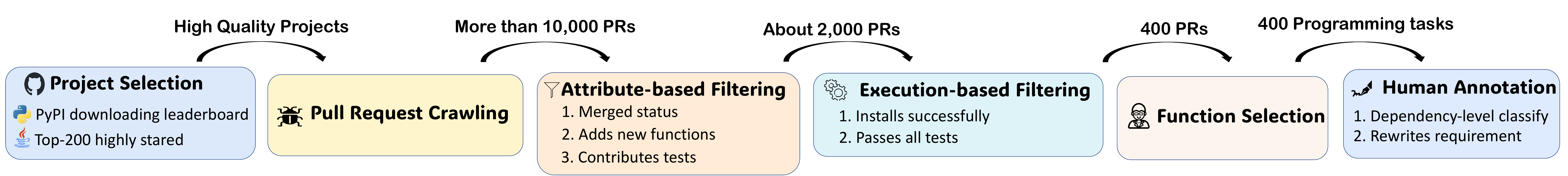}
  \caption{\HumanEvo construction pipeline.}
  \label{fig:pipeline}
\end{figure*}

\ylrevised{
As depicted in Figure~\ref{fig:pipeline}, the construction pipeline of \HumanEvo consists of six steps, namely, (1) Project Selection; (2) Pull Request Crawling; (3) Attribute-based Filtering; (4) Execution-based Filtering; (5) Function Selection; and (6) Human Annotation. 
}

\textbf{(1) Project Selection.}
In order to make \HumanEvo more representative, we adopt a strict approach to selecting functions from various open-source projects. For Python, we select high-quality real-world projects from a list of the top 5,000 most downloaded PyPI libraries\footnote{\url{https://hugovk.github.io/top-pypi-packages/}}. Subsequently, we deploy the selected projects and run their testing frameworks to ensure each of them can run correctly and stably. We then proceed to collect PRs from these repositories through the GitHub developer API\footnote{\url{https://api.github.com}}. 
For Java, as widely recognized projects typically indicate extensive documentation, structured open-source development guidelines, and functional, well-formatted code, our approach involves searching for the top 200 starred projects on GitHub. Then, we deploy every project on the list to select projects that can be successfully deployed and pass the corresponding test suite. \dewu{Finally, we retain 30 real-world projects, covering over 50 common development domains, which guarantees that the functions selected for evaluation in \HumanEvo could represent real-world programming scenarios.}
 
\textbf{(2) Pull Request Crawling.}
We crawl a large number of initial PRs from the popular open-source repositories selected in the procedure above. Specifically, \dewu{for each project, we do not crawl all PRs as we find that early versions of many projects are difficult to deploy successfully. Therefore, for Python, we crawl PRs created after September 2015 (the release time of Python 3.5); for Java, we collect PRs created after September 2014 (the release time of Java 8). Finally, }we obtain over 10,000 PRs and meticulously record metadata pertaining to these PRs for subsequent filtering processes.

\textbf{(3) Attribute-based Filtering.}
In this step, we preliminarily filter the crawled PRs based on the following three attributes.
\begin{itemize}
    \item The status of the PR is ``Merged''. A ``Merged'' status indicates that the code changes associated with the PR were accepted and incorporated into its parent repository.
    \item The PR must introduce new functions. These newly added functions are typically intended to develop the software project by adding new features, improvements, or fulfilling general development objectives, which can be selected as programming task.
    \item The PR must introduce at least one or more new tests, and these tests must cover the selected function. 
\end{itemize}

All these attributes aim to ensure that the subsequently selected functions are of high quality, have undergone rigorous review and are covered by the project's testing framework.

\textbf{(4) Execution-based Filtering.}
In this step, we perform execution-based filtering for the PRs filtered in the attributes filtering step. \dewu{Note that, although both the project selection step and the execution-based filtering step have the "Successful installation" criterion, their purposes are different. The main purpose of requiring successful installation in the project selection step is to quickly filter projects, while in this step, it ensures that all versions associated with different PRs can run successfully, as PRs may correspond to multiple versions of a project.}

Firstly, we prepare the execution environment for each candidate task instance in \HumanEvo. For each task instance, we first determine the version of the project corresponding to the PR, then manually identify the required dependency environment for running the project at that version. For example, for a Python task instance, we confirm the required Python version and the types and versions of third-party libraries it depends on. After obtaining this dependency environment information, we create a virtual execution environment for each version of the project. Task instances that can not install the dependency environment properly are filtered out during \dewu{the virtual execution environment setup process.}
After creating the execution environment, we execute the project's testing framework in the corresponding environment.

Secondly, for each selected function, we execute the project's test framework in its required environment to ensure that the selected function is covered by the project's test framework. Specifically, we split the PR's patch file, which is also known as the code diff file, into two parts, namely the production code patch and the test code patch. For each candidate task, we apply the PR’s test patch and log the associated test results before and after applying the PR’s production patch. Then, we filter out task instances with no test where its status changes from a fail to pass, which indicates the newly added functions are covered by the test suite. After filtering out instances without a fail-to-pass transition, we obtain 400 PRs.

\textbf{(5) Function Selection.}
In this step, we manually select functions from the PRs that have passed the filtering process above to serve as programming tasks. After completing the above filtering, we confirm that the remaining PRs meet our requirements. Therefore, we select newly added functions from these PRs as programming tasks. During the selection process, to avoid choosing basic functions, we filter out initialization functions, constructors, and destructors, to ensure the quality of the selected programming tasks. \dewu{In addition, we ensure that the selected target functions are covered by the project's test framework through manual review .} Finally, we acquire a total of 400 programming tasks.

\textbf{(6) Human Annotation.}
In this step, we recruit six annotators with more than five years of programming experience to manually rewrite two different styles of docstrings for the selected functions and label each function with a dependency level. 
We rewrite the docstring for each instance in \HumanEvo to ensure the quality of the benchmark dataset and mitigate data leakage issues as much as possible. 
Specifically, in the process of data construction, we find that the docstring styles of actual projects can be divided into two categories: the detail and the brief.
\dewu{Docstrings are used as input for models, so variations in docstring style may significantly influence model evaluation. Aiming to determine whether the evolution-ignored setting leads to unreal LLMs' performance evaluations across different input styles,} we provide two types of the docstring for each task instance. The detailed docstring describes the function of the target function in detail, as well as the types and meanings of its inputs and outputs. In contrast, a brief docstring only briefly describes what the target function does. After the annotators complete writing the docstrings, we conduct a cross-validation review, requiring each task instance to be verified and confirmed by two developers to ensure accuracy. If the annotators disagree with a docstring written by the previous annotator during the review process, they will discuss the issue together until all annotators reach an agreement.

In order to facilitate the evaluation of the ability of LLMs in generating functions with different dependency levels, we further categorize all task instances in \HumanEvo into three categories: standalone, intra-class, and inter-class functions. A dependency denotes the invocation of elements defined within projects, and the dependency level is defined based on the position of the target function's dependencies. Standalone functions refer to those that invoke or access only built-in functions and standard libraries. Intra-class functions rely on other functions within the same class. Inter-class functions involve calling functions within the same file or functions in other files. 




\subsection{Benchmark Characteristics}
\label{sec:benchmark_characteristics}

\subsubsection{\textbf{Evolution-aware Benchmark}} \HumanEvo is the pioneering repository-level code generation benchmark to introduce the evolution-aware concept. We highlight a significant oversight in prior research: the developmental trajectory of a project is an evolution process, altering the project context available to programmers as the project advances through various stages. Previous studies have overlooked this temporal dimension, resulting in LLMs being fed with code that had not yet been implemented in the project when certain functions were developed. Consequently, it also misses some context that should have been present but might have been deleted or updated during the project's evolution. This undoubtedly leads to the model facing an unrealistic development scenario. As a result, the performance of LLMs in real-world software development tasks remains unknown. Therefore, we are committed to enhancing \HumanEvo to serve as a benchmark that better simulates real-world software code development scenarios, aiming to more accurately reflect the true performance of LLMs applied to pragmatic development tasks.


\subsubsection{\textbf{Strict Data Filtering Process}} We ensure the quality of the selected function from several aspects. First, we ensure that the selected projects are of high quality because high-quality projects generally have excellent maintenance, comprehensive testing frameworks, and good coding styles. For Python, we choose the projects with the highest download count in PyPI. For Java, we select projects from the top 200 projects on GitHub based on the number of stars, with an average of around 24k stars. Previous benchmarks have ensured the functionality of target functions by writing unit tests for each function separately. However, we believe that in a project with complex dependencies, adding a new function may have certain impacts on other parts of the project. Relying solely on unit tests for the target function is insufficient to ensure that the project can still function properly and pass the project's test framework after inserting the function. Therefore, in this work, we directly run the project's test framework to ensure the correctness of the target function's functionality while ensuring that inserting the generated function does not have a negative impact on other parts of the project. 

\dewu{In addition, we also verify that our selected functions are covered by the project's test framework through execution-based validation. First, we decompose the code diff in the PR corresponding to the target function into a source code patch and a test code patch, and directly apply the test code patch. Then, we validate whether the target function is covered by the project's test framework by executing the project's test framework twice. In the first execution, we run the project's test framework without applying the source code patch. In the second execution, we apply the source code patch and then run the test framework. If there is a test function in the project's test framework that targets the target function, and the execution of this test function requires calling the target function, the first execution (based on execution) will fail because the target function call is missing. In the second execution, after applying the source code patch, the test framework will correctly call the target function and run as expected. Therefore, if the test framework covers the target function, the results of the two executions should show a ``fail-to-pass'' transition. We perform execution-based validation for all selected functions to ensure that the \HumanEvo test suite is reliable.}

\begin{table}[t]
\centering
\caption{Dependency level in \HumanEvo.}
\label{dependency level}
\begin{tabular}{ccc}
\toprule
Dependency Level& \HumanEvo-Python &\HumanEvo-Java\\
\midrule
Standalone & \dewu{45 (22.5\%)} &26 (13.0\%)\\
Intra-class& \dewu{60 (30.0\%)} &61 (30.5\%)\\
Inter-class & \dewu{95 (47.5\%)} & 113 (56.5\%) \\
\bottomrule
\end{tabular}
\end{table}
\subsubsection{\textbf{Human Annotation}}

We manually rewrite two styles of docstrings for all task instances and label each of them with a dependency level. During the dataset construction process, we observe different docstring styles among these high-quality projects, which can be generally categorized into the detailed and the brief. Detailed docstring typically provides comprehensive descriptions of a function's functionality, elucidates input and output parameter types, and annotates potential error scenarios. Conversely, a brief docstring succinctly outlines the function's functionality in a few sentences. To enhance the diversity of our benchmark and better align with real-world development scenarios, we manually crafted two types of docstring for each function: detailed and brief. This provides users with more options for subsequent usage.

Furthermore, to facilitate the assessment of LLMs' capability in generating functions with varying levels of dependency, we categorize the selected functions based on their dependency levels, including standalone, intra-class, and inter-class. Standalone functions can be implemented without relying on other functions within the project. Intra-class functions require dependencies on other functions within the same class. Inter-class functions imply more complex dependencies, with most of the dependent functions being located in other files within the project. Manual categorization of dependency levels aids in analyzing the performance of LLMs in generating functions of varying complexity, facilitating researchers in improving the application of LLMs in practical development scenarios based on different contexts. Moreover, as indicated in Table~\ref{dependency level}, the majority of our benchmark consists of functions with deep dependency levels. This suggests that our benchmark presents a sufficiently challenging scenario.

Note that we use PRs to construct \HumanEvo primarily for the following three reasons. First, since prompting LLMs to generate repository-level functions can be viewed as an incremental development process, and PRs for adding new features exactly mirror the incremental development process in the real world, we choose to instruct LLMs generate the newly added functions in PRs to simulate real-world development scenarios. Second, the quality of functions in PRs is guaranteed. Before merging the newly added code in PRs into the project, repository administrators conduct a review on it, which allows us to select high-quality task instances directly. Third, the base commit in PRs helps us roll back the project to the state before the target function was committed. Finally, by processing PRs appropriately, we can further determine whether the target functions are covered by the project's testing framework.



\section{Experimental Design}


In this section, we introduce the models used in the experiments, outline the research questions that the experiments aim to address and provide a detailed description of the experimental setups. 

\subsection{Studied LLMs}

\ylrevised{
We select the mainstream LLMs (both open-source and closed-source ones) that have been widely used in recent code generation work~\cite{phan2024repohyper,codereval,evocodebench,repocoder}. 
For open-source LLMs, we select the CodeLlama series~\cite{codellama} (including {CodeLlama-7B}, {CodeLlama-13B}, and {CodeLlama-34B}) and the DeepSeekCoder series~\cite{deepseek} (including {DeepSeekCoder-6.7B} and {DeepSeekCoder-33B}). The CodeLlama and DeepSeekCoder series are popular LLMs that perform well in code generation. 
For closed-source LLMs, we choose the commonly used commercial models: {GPT-3.5-Turbo}~\cite{gpt3.5} and {GPT-4}~\cite{gpt4}, with GPT-4 demonstrating excellent performance in previous benchmarks.
}

\subsection{Studied Context Acquisition Strategies}
Due to the large size of modern software repositories, it is not feasible to feed the entire repository to the model. Thus, previous works mainly obtain project context that might be beneficial for generating target functions through two strategies: retrieval-based and static analysis-based methods. In the following experiments, we cover both of these technical routes to demonstrate that the evolution-ignored situation leads to the deviation of the LLMs' performance from reality.

\textbf{Retrieval-based Context Acquisition.}
We adopt the standard retrieval-augmented-generation pipeline for repository-level code generation~\cite{repocoder, distillation, colbertv2,approximate}. In the evolution-aware setting, we roll back the repository to the state before the target code was committed for each task instance. On the contrary, in the evolution-ignored settings, we directly utilize the latest project version as the context source. Then, we construct a corresponding context retrieval database by partitioning the source code files from the repository into a collection of code snippets. Subsequently, we utilize the docstring of the target code as the query for retrieval and compute the Jaccard similarity between the query and all code snippets in the context retrieval database. After sorting all the code snippets by similarity score, we select project contexts with higher similarity scores as project context for target function generation.

\textbf{Static Analysis-based Context Acquisition.}
Similar to the static analysis-based repository-level code generation work~\cite{repositorylevel}, we obtain code files in the repository that might help generate the target code through static analysis. First, we roll back the repository to the corresponding state according to evolution-ignored or evolution-aware settings. Then, we extract the source code from these files as project context to feed to the model. There are four types of context sources: local file (the file where the target code resides), import file (files imported by the local file in the project), sibling file (files located in the same folder as the local file), and similar file (files in the project with names similar to the local file). 
We utilize four prompting strategies from ~\cite{repositorylevel}: \textbf{local} – extracting context solely from the local file; \textbf{local+import} – extracting code from both local and import files; \textbf{local+sibling} - extracting code from both local and sibling files; \textbf{local+sim} - extracting code from both local and similar files.


\subsection{Research Questions}
Our experiments intend to answer the following research questions (RQs):
\begin{itemize}
    \item \textbf{RQ1: How effective is \HumanEvo in benchmarking the performance of LLMs in repository-level code generation task?}

    We conduct experiments under two different settings, evolution-aware and evolution-ignored, to explore the differences in model performance between these two settings and reveal the actual performance of the models in real development scenarios.
    
    \item \textbf{RQ2: How does the code generation performance change as repository evolves?}

     We treat multiple versions of project as context sources to explore how the model's performance change as the project evolves.
    \item \textbf{RQ3: How different code descriptions in the dataset affect code generation performance?}
 
    We conduct experiments to compare the performance of the model under different docstring styles.
\end{itemize}

\subsection{Experimental Settings}
We assess the functionality correctness of code generated by LLMs with execution-based evaluation and utilize the commonly used metric {Pass@1} for evaluation. \dewu{We set the generation temperature of LLMs to 0.8, top-p to 0.95, and the context window length to 4096.} Note that, to mitigate issues stemming from the randomness of model generation, the experimental results presented in this paper are obtained by conducting three repeated experiments and averaging the results. The experiment was conducted on a server running Ubuntu 18.04.6 LTS and equipped with 128 Intel(R) Xeon(R) Platinum 8336C @ 2.30GHzE CPUs, and 8 NVIDIA A800 with 80GB memory.


\begin{table}[t]
    \centering
    \setlength{\tabcolsep}{5pt}
    \caption{Retrieval-based context acquisition approach: model performance under evolution-aware (\HumanEvo) and evolution-ignored (EI) settings. } 
    \label{LLM's performance on retrieval based approach}
    \scalebox{0.9}{
    \begin{tabular}{l cc cc}
    \toprule
        \multirow{3}{*}{Model}  & \multicolumn{2}{c}{Python} &  \multicolumn{2}{c}{Java}\\ 
        \cmidrule(lr){2-3} \cmidrule(lr){4-5}
       ~ & EI & \HumanEvo &EI & \HumanEvo \\
       \midrule
       CodeLlama-7B  &29.0\%&\cellcolor{blue!5}22.0\% ($\downarrow 24.1\%$)&8.5\%&\cellcolor{blue!5}6.0\% ($\downarrow 29.4\%$)\\
       CodeLlama-13B &30.5\%&\cellcolor{blue!5}23.0\% ($\downarrow 24.6\%$)&9.5\%&\cellcolor{blue!5}7.0\% ($\downarrow 26.3\%$)\\ 
       CodeLlama-34B &31.5\%&\cellcolor{blue!5}24.0\% ($\downarrow 23.8\%$)&14.5\%&\cellcolor{blue!5}10.0\% ($\downarrow 31.0\%$)\\
       DeepSeekCoder-6.7B &30.0\%&\cellcolor{blue!5}23.0\% ($\downarrow 23.3\%$)&13.0\%&\cellcolor{blue!5}9.0\% ($\downarrow 30.8\%$)\\
      DeepSeekCoder-33B &31.0\%&\cellcolor{blue!5}25.0\% ($\downarrow 19.4\%$)&15.5\%&\cellcolor{blue!5}11.5\%( $\downarrow 25.8\%$)\\
       GPT-3.5-Turbo &32.5\%&\cellcolor{blue!5}25.5\% ($\downarrow 21.5\%$)&18.0\%&\cellcolor{blue!5}13.0\% ($\downarrow 27.8\%$)\\
       GPT-4&34.5\%&\cellcolor{blue!5}26.5\% ($\downarrow 23.2\%$)&20.5\%&\cellcolor{blue!5}14.5\% ($\downarrow 29.3\%$)\\\bottomrule
    \end{tabular}
    }
\end{table}
\section{Evaluation Results}

\begin{table*}[t]
    \centering
    \setlength{\tabcolsep}{3.5pt}
    \caption{Static analysis based project context acquisition approaches: model performance under evolution-aware (\HumanEvo) and evolution-ignored (EI) settings.}
    \label{LLM's performance on static anaysis based approach}
    {\begin{tabular}{l l cc cc cc cc}
    \toprule
    \multirow{2}{*}{Data} &\multirow{2}{*}{Model} 
     & \multicolumn{2}{c}{local} & \multicolumn{2}{c}{local+import}&\multicolumn{2}{c}{local+sibling} &\multicolumn{2}{c}{local+sim}\\
     \cmidrule(lr){3-4} \cmidrule(lr){5-6} \cmidrule(lr){7-8} \cmidrule(lr){9-10}
     ~ &~& EI & \HumanEvo &EI & \HumanEvo &EI & \HumanEvo &EI & \HumanEvo\\ \midrule
     \multirow{5}{*}{\HumanEvo-Python} &CodeLlama-7B  
       &27.0\%& \cellcolor{blue!5}24.0\%($\downarrow 11.1\%$)
       &28.0\%& \cellcolor{blue!5}25.0\%($\downarrow 10.7\%$)
       &28.5\%& \cellcolor{blue!5}25.0\%($\downarrow 12.3\%$)
       &29.0\%& \cellcolor{blue!5}25.0\%($\downarrow 13.8\%$)\\
       ~ &CodeLlama-13B 
       &28.5\%& \cellcolor{blue!5}24.5\%($\downarrow 14.0\%$)
       &29.5\%& \cellcolor{blue!5}26.5\%($\downarrow 10.2\%$)
       &29.5\%& \cellcolor{blue!5}25.5\%($\downarrow 13.6\%$)
       &30.5\%& \cellcolor{blue!5}25.5\%($\downarrow 16.4\%$)\\
       ~ &CodeLlama-34B 
       &29.5\%& \cellcolor{blue!5}25.0\%($\downarrow 15.3\%$) 
       &30.0\%& \cellcolor{blue!5}27.0\%($\downarrow 10.0\%$)  
       &30.0\%& \cellcolor{blue!5}26.5\%($\downarrow 11.7\%$)  
       &30.5\%& \cellcolor{blue!5}26.5\%($\downarrow 13.1\%$)\\
       ~&DeepSeekCoder-6.7B 
       &29.0\%& \cellcolor{blue!5}24.0\%($\downarrow 17.2\%$)  
       &29.0\%& \cellcolor{blue!5}24.5\%($\downarrow 15.5\%$)  
       &29.0\%& \cellcolor{blue!5}25.5\%($\downarrow 12.1\%$)   
       &29.5\%& \cellcolor{blue!5}26.0\%($\downarrow 12.1\%$)\\
       ~&DeepSeekCoder-33B 
       &30.5\%& \cellcolor{blue!5}25.5\%($\downarrow 16.4\%$)  
       &31.0\%& \cellcolor{blue!5}26.5\%($\downarrow 14.5\%$)  
       &30.5\%& \cellcolor{blue!5}26.5\%($\downarrow 13.1\%$)
       &31.0\%& \cellcolor{blue!5}26.5\%($\downarrow 14.5\%$)\\
       ~&GPT-3.5-Turbo
       &33.0\%& \cellcolor{blue!5}26.5\%($\downarrow 19.7\%$)  
       &34.0\%& \cellcolor{blue!5}27.0\%($\downarrow 20.6\%$)  
       &32.5\%& \cellcolor{blue!5}26.5\%($\downarrow 18.5\%$)  
       &33.5\%& \cellcolor{blue!5}27.5\%($\downarrow 17.9\%$)
        \\
       ~&GPT-4
       &34.5\%& \cellcolor{blue!5}27.0\%($\downarrow 21.7\%$)  
       &35.0\%& \cellcolor{blue!5}28.0\%($\downarrow 20.0\%$)  
       &34.5\%& \cellcolor{blue!5}27.5\%($\downarrow 20.3\%$)  
       &35.5\%& \cellcolor{blue!5}28.0\%($\downarrow 21.1\%$)
       \\
       \midrule
       \multirow{5}{*}{\HumanEvo-Java} 
       &CodeLlama-7B
       &16.5\%& \cellcolor{blue!5}8.0\%($\downarrow 51.5\%$)
       &15.0\%& \cellcolor{blue!5}6.5\%($\downarrow 56.7\%$)
       &12.0\%& \cellcolor{blue!5}6.0\%($\downarrow 50.0\%$)
       &13.5\%& \cellcolor{blue!5}6.5\%($\downarrow 51.9\%$) \\
       ~&CodeLlama-13B
       &15.0\%& \cellcolor{blue!5}6.5\%($\downarrow 56.7\%$)   
       &16.0\%& \cellcolor{blue!5}7.0\%($\downarrow 56.3\%$)  
       &15.5\%& \cellcolor{blue!5}6.5\%($\downarrow 58.1\%$) 
       &15.5\%& \cellcolor{blue!5}6.5\%($\downarrow 58.1\%$) \\
       ~&CodeLlama-34B
       &17.5\%& \cellcolor{blue!5}8.0\%($\downarrow 54.3\%$)  
       &14.5\%& \cellcolor{blue!5}7.0\%($\downarrow 51.7\%$)   
       &12.5\%& \cellcolor{blue!5}6.5\%($\downarrow 48.0\%$)  
       &14.5\%& \cellcolor{blue!5}8.0\%($\downarrow 44.8\%$)  \\
       ~&DeepSeekCoder-6.7B 
       &17.0\%& \cellcolor{blue!5}8.5\%($\downarrow 50.0\%$)  
       &16.5\%& \cellcolor{blue!5}10.0\%($\downarrow 39.3\%$)  
       &14.5\%& \cellcolor{blue!5}9.0\%($\downarrow 37.9\%$)   
       &18.0\%& \cellcolor{blue!5}7.0\%($\downarrow 61.1\%$)  \\
       ~&DeepSeekCoder-33B
       &16.5\%& \cellcolor{blue!5}10.0\%($\downarrow 39.3\%$)
       &16.5\%& \cellcolor{blue!5}10.0\%($\downarrow 39.3\%$)  
       &18.5\%& \cellcolor{blue!5}8.5\%($\downarrow 54.1\%$)    
       &19.0\%& \cellcolor{blue!5}10.0\%($\downarrow  47.4\%$) \\
       ~&GPT-3.5-Turbo
       &19.5\%& \cellcolor{blue!5}11.5\%($\downarrow 41.0\%$)
       &21.0\%& \cellcolor{blue!5}14.0\%($\downarrow 33.3\%$)
       &20.0\%& \cellcolor{blue!5}12.0\%($\downarrow 40.0\%$)
       &21.5\%& \cellcolor{blue!5}13.0\%($\downarrow 39.5\%$)
       \\
       ~&GPT-4  
       &22.0\%& \cellcolor{blue!5}13.5\%($\downarrow 38.6\%$)
       &23.0\%& \cellcolor{blue!5}14.5\%($\downarrow 37.0\%$)
       &22.5\%& \cellcolor{blue!5}14.0\%($\downarrow 37.8\%$)
       &23.5\%& \cellcolor{blue!5}14.5\%($\downarrow 38.3\%$)
       \\
       \bottomrule
    \end{tabular}
    }
\end{table*}

\subsection{RQ1: How effective is \HumanEvo in benchmarking the performance of LLMs in repository-level code generation task?}

\textbf{Overall Performance.}
Table~\ref{LLM's performance on retrieval based approach} and Table~\ref{LLM's performance on static anaysis based approach} present the performance of seven mainstream LLMs under two different experimental settings: \emph{Evolution-Ignored} (where the context source is the latest version of the project) and \emph{Evolution-Aware} (\HumanEvo, where the context source is the project before the target function was committed). Table~\ref{LLM's performance on retrieval based approach} shows the performance of retrieval based project context acquisition approach, and Table~\ref{LLM's performance on static anaysis based approach} shows the performance of static analysis based project context acquisition approaches.
All the experimental results in Table~\ref{LLM's performance on retrieval based approach} and Table~\ref{LLM's performance on static anaysis based approach} show a significant decrease in performance under the evolution-aware setting compared to the evolution-ignored setting. \dewu{This suggests that ignoring the evolution of the project would lead to inflated performance of LLMs, failing to accurately reflect their real code generation performance.}

\begin{table}[t]
    \centering
    \caption{Break down analysis. EA and EI stand for evolution-aware and evolution-ignored, respectively. \checkmark means success, \scalebox{0.85}[1]{$\times$} meas failure.}
    \setlength{\tabcolsep}{4pt}
    \label{overlap}
    \begin{tabular}{l ccc ccc}
    \toprule
        \multirow{3}{*}{Model}  & \multicolumn{3}{c}{Python} &  \multicolumn{3}{c}{Java}\\ 
        \cmidrule(lr){2-4} \cmidrule(lr){5-7}
       ~ & EA\checkmark  & EA\checkmark  &EA\scalebox{0.85}[1]{$\times$}    & EA\checkmark  & EA\checkmark  &EA\scalebox{0.85}[1]{$\times$}   \\
        ~ &EI\scalebox{0.85}[1]{$\times$}  &EI\checkmark  & EI\checkmark &EI\scalebox{0.85}[1]{$\times$}    &EI\checkmark  &EI\checkmark

       \\
       \midrule
       CodeLlama-7B  &7&37&21  &3&9&8\\
       CodeLlama-13B &4&42&19  &3&11&8\\ 
       CodeLlama-34B &2&46&17  &6&14&15\\
       DeepSeekCoder-6.7B &4&42&18  &7&11&15\\
       DeepSeekCoder-33B &4&46&16   &5&18&13\\
       GPT-3.5-Turbo &2&49&16   &10&16&20\\
       GPT-4         &0&53&15       &4&25&16\\\bottomrule
    \end{tabular}
\end{table}

\begin{table}[t]
    \centering \small
    \caption{Model performance in generating functions with deeper dependency levels.}
    \label{dependency}
    \setlength{\tabcolsep}{2pt}
    \scalebox{0.8}{
    \begin{tabular}{c l cc cc}
    \toprule
        \multirow{3}{*}{} &\multirow{3}{*}{Model}  & \multicolumn{2}{c}{Python} &  \multicolumn{2}{c}{Java}\\ 
        \cmidrule(lr){3-4} \cmidrule(lr){5-6}
       ~& & EI & \HumanEvo  & EI & \HumanEvo \\\midrule
       \multirow{7}{*}{Intra-Class} 
         &CodeLlama-7B &32.8\% &\cellcolor{blue!5}31.1\% ($\downarrow 5.0\%$)&1.6\%&\cellcolor{blue!5}1.6\%\\
       ~ &CodeLlama-13B &29.5\%&\cellcolor{blue!5}27.9\% ($\downarrow 5.6\%$)&2.9\%&\cellcolor{blue!5}1.9\% ($\downarrow 33.3\%$)\\
       ~ &CodeLlama-34B &34.4\%&\cellcolor{blue!5}27.9\% ($\downarrow 19.0\%$)&8.2\%&\cellcolor{blue!5}6.6\% ($\downarrow 20.0\%$)\\
       ~ &DeepSeekCoder-6.7B  &34.4\%&\cellcolor{blue!5}27.9\% ($\downarrow 19.0\%$)&4.9\%&\cellcolor{blue!5}3.3\% ($\downarrow 33.3\%$)\\
       ~ &DeepSeekCoder-33B &37.7\% &\cellcolor{blue!5}34.4\% ($\downarrow 8.7\%$)&3.3\%&\cellcolor{blue!5}3.3\%\\
       ~ &GPT-3.5-Turbo   &36.1\%&\cellcolor{blue!5}29.5\% ($\downarrow 18.2\%$)&4.9\%&\cellcolor{blue!5}4.9\%\\
       ~ &GPT-4    &37.7\%&\cellcolor{blue!5}32.9\% ($\downarrow 13.0\%$)&8.2\%&\cellcolor{blue!5}4.9\% ($\downarrow 40.0\%$)\\
       \midrule
       \multirow{7}{*}{Inter-Class} 
         &CodeLlama-7B &27.2\%&\cellcolor{blue!5}17.5\% ($\downarrow 35.7\%$)&10.6\%&\cellcolor{blue!5}5.3\% ($\downarrow 50.0\%$)\\
       ~ &CodeLlama-13B &31.1\%&\cellcolor{blue!5}20.3\% ($\downarrow 34.4\%$)&10.6\%&\cellcolor{blue!5}6.2\% ($\downarrow 41.7\%$)\\
       ~ &CodeLlama-34B &30.1\%&\cellcolor{blue!5}22.3\% ($\downarrow 25.8\%$)&15.0\%&\cellcolor{blue!5}10.6\% ($\downarrow 29.4\%$)\\
       ~ &DeepSeekCoder-6.7B  &27.2\%&\cellcolor{blue!5}20.4\% ($\downarrow 25.0\%$)&15.9\%&\cellcolor{blue!5}11.5\% ($\downarrow 27.8\%$)\\
       ~ &DeepSeekCoder-33B &28.2\%&\cellcolor{blue!5}21.4\% ($\downarrow 24.1\%$)&17.7\%&\cellcolor{blue!5}13.3\% ($\downarrow 25\%$)\\
       ~ &GPT-3.5-Turbo   &31.1\%&\cellcolor{blue!5}23.3\% ($\downarrow 25.0\%$)&21.2\%&\cellcolor{blue!5}14.2\% ($\downarrow 33.3\%$)\\
       ~ &GPT-4    &32.0\%&\cellcolor{blue!5}23.3\% ($\downarrow 27.3\%$)&22.1\%&\cellcolor{blue!5}15.9\% ($\downarrow 28.0\%$)\\\bottomrule
       
    \end{tabular}
    }
\end{table}

\begin{figure*}[t]
  \centering
  \includegraphics[width=0.85\textwidth]{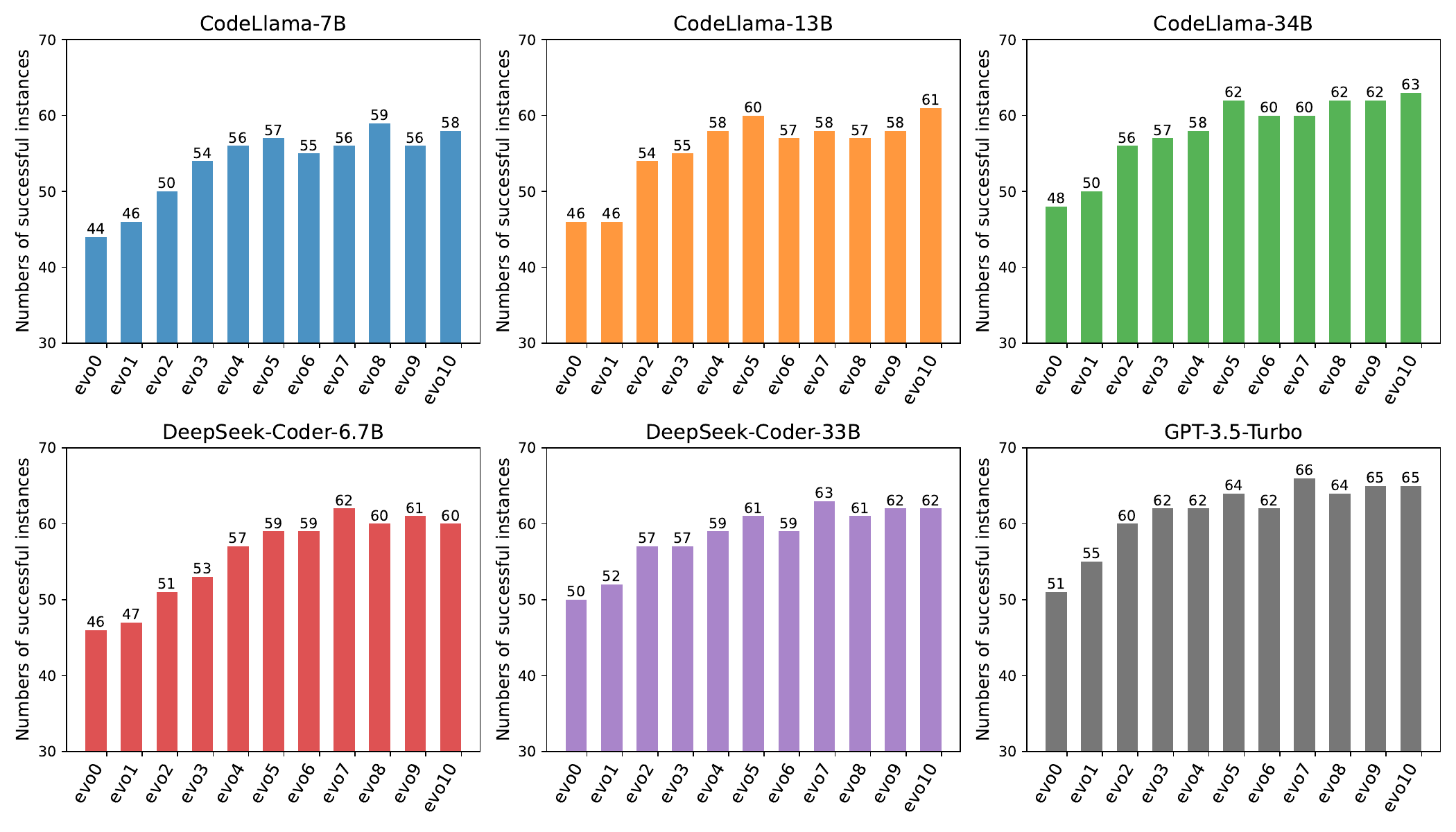}
  \caption{LLM's performance when using different project versions as context sources.}
  \label{RQ2}
\end{figure*}
\textbf{Break Down Analysis.}
We further analyze the generation results to investigate whether evolution-ignored setting always leads to over-estimated performance of LLMs. 
Table~\ref{overlap} illustrates the overlap of successful task instances between evolution-aware setting and evolution-ignored setting. Since the results from retrieval-based project context acquisition approach and static analysis based project context acquisition approaches are similar, subsequent analysis primarily focuses on the experimental results of the retrieval-based approach. As introduced above, \dewu{evolution-ignored setting would raise issues such as }the future context leakage issue\dewu{, which} would lead to inflated performance.   \dewu{Meanwhile, evolution-ignored setting can raise issues such as }the useful context leakage issue\dewu{, which} would lead to underestimated performance. However, the results in Table~\ref{LLM's performance on retrieval based approach} reflect that models' overall performance only exhibits an inflated trend, driving us to delve deeper into whether 
\dewu{issues such as }useful context leakage has occurred. We investigate the overlap between task instances that LLMs successfully completed under evolution-ignored and evolution-aware settings.

In Table~\ref{overlap}, EA stands for the evolution-aware setting and EI stands for the evolution-ignored setting. A \checkmark means successful, while a \scalebox{0.85}[1]{$\times$} means failure. So, EA\checkmark EI\scalebox{0.85}[1]{$\times$} refers to the number of instances that pass under the evolution-aware setting but fail under the evolution-ignored setting. EA\scalebox{0.85}[1]{$\times$} EI\checkmark represents instances that fail under EA but pass under EI. EA\checkmark EI\checkmark denotes the number of instances that pass under both settings. From the experimental results, it is evident that except for GPT-4 in \HumanEvo-Python, which has a value of 0 for EA\checkmark EI\scalebox{0.85}[1]{$\times$} situation, all other LLMs exhibit deviations from reality caused by \dewu{issues such as the useful context missing issue} in a certain quantity, showing that the issues\dewu{, which are raised by the evolution-ignored setting and would lead to underestimated performance,}
have impact on evolution-ignored setting. 
However, due to the significantly higher number in EA\scalebox{0.85}[1]{$\times$} EI\checkmark, the primary influencing factor remains the future context leakage, leading to exaggerated performance of LLMs.

\textbf{Dependency Level.}
Functions with different dependency levels vary in their reliance on project context, and the evolution-ignored setting directly affects project context. Therefore, in this experiment, We examine the performance differences of LLMs in generating functions with deeper dependency levels, focusing on non-standalone functions in \HumanEvo, which encompass intra-class and inter-class functions, across two settings: evolution-ignored and evolution-aware.
As shown in Table~\ref{dependency}, under the evolution-aware setting, the accuracy of LLM in generating both intra-class and inter-class functions has noticeably declined. The decline is more pronounced for inter-class functions, as these functions typically require cross-file context, which is often more scattered. As the project evolves, its various parts tend to become more comprehensive. Therefore, under the evolution-ignored setting, the models can reference an increasing amount of context. Consequently, when the evolution-aware setting loses access to these contexts, the model's performance significantly deteriorates.


\begin{center}
    \begin{myboxc}{\textbf{RQ1 Summary: }
    Experiments show an inflated trend in LLM's performance under the evolution-ignored setting. 
    }
    \end{myboxc}
\end{center}


\subsection{RQ2: How does the code generation performance change as repository evolves?}

To further demonstrate the impact of evolution-ignored setting, we treat different versions of the project as context sources to explore how the performance of LLMs change over time. 
We start from the version right before the target code was committed and incrementally increase the project versions up to the latest version. 
Typically, a project's version number consists of two digits in the format ``v1.2'', where the first digit denotes the major version and the second digit denotes the minor version.  
In Figure~\ref{RQ2}, evo0 is the version right before the target code was committed, standing for the evolution-aware setting. 
For evo1 to evo10, the version numbers are incrementally increased. For example, if evo0 corresponds to the ``v1.2'' version of the project, then evo1 will be ``v1.3''.


As shown in Figure~\ref{RQ2}, we find that the performance of LLMs deviates from the actual performance in the evo0 (evolution-aware) setting, with this deviation increasing as the project evolves. 
All studied LLMs, regardless of being open-source or closed-source and irrespective of parameter size, consistently exhibit inflated performance and follow a similar trend in performance changes. 
This result further confirms that ignoring project evolution can lead to unrealistic evaluation of LLMs. 


\begin{center}
    \begin{myboxc}{\textbf{RQ2 Summary: }
    In the evolution-ignored setting, as the project evolves, the performance of LLMs deviates from actual performance, with this deviation increasing over time.
    } 
    \end{myboxc}
\end{center}

\subsection{RQ3: How different code descriptions in the dataset affect code generation performance?}

During the benchmark construction process, we notice that, in real-world development, some projects tend to write detailed docstrings, while the others prefer to describe the function's functionality using only several brief statements. 
\dewu{Since these docstrings are used as input to models which may significantly affect model performance, we explore whether the impact of the evolution-ignored setting on model performance varies under different docstring styles. To this end,}
we manually write both brief and detailed docstrings for each task instance in \HumanEvo. In this experiment, we evaluate the performance of the studied LLMs on the two types of manually written docstrings. As shown in Table~\ref{RQ3}, experimental results indicate that \dewu{under the evolution-ignored setting, using different styles of docstrings as input consistently leads to inflated performance, and the style of the docstring does not significantly change the impact of the evolution-ignored setting on the inflated performance evaluation. Additionally,} the majority of LLMs perform better on detailed docstrings compared to the brief docstrings. We suspect that the detailed docstrings provide a more comprehensive description of the functionality of the target function and include the implementation logic of the target function. In addition, detailed docstrings also provide explanations for the input and output of the target function, enabling LLMs to better understand the tasks to be addressed. Conversely, brief docstrings only introduce the intention of the target function, which may lead to LLMs struggling to understand user intent in certain complex scenarios, ultimately resulting in decreased performance.

However, the decline in LLMs' performance is not significant. We suspect that the reason could possibly be related to our use of a retrieval-augmented generation approach, by which we have retrieved a substantial amount of context from the project through similarity calculations. These contexts contribute to a lengthy prompt. However, when LLMs deal with overly long prompts, their attention gets dispersed, leading to the neglect of some useful information by the model.

\begin{center}
    \begin{myboxc}{\textbf{RQ3 Summary: }
   \dewu{From the experimental results, we do not observe any correlation between docstring styles and evolution settings.} Compared to brief docstrings, detailed docstrings provide more beneficial information to LLMs, resulting in better performance. 
   However, the performance improvement on detailed docstrings is not substantial.
    }
    \end{myboxc}
\end{center}

\section{Threats to Validity}

We have identified the following potential threats that may affect the validity of our study. 
Firstly, due to limited computational resources, our experiments were not able to cover all the available code LLMs, thus not fully reflecting the performance of all LLMs in actual development scenarios, which may slightly affect the representativeness of our experiments. 
Secondly, currently \HumanEvo only includes two programming languages. In the future, we plan to address this limitation by continuously expanding \HumanEvo to cover as many programming languages as possible, thereby providing a more comprehensive evaluation.
Thirdly, in prompt format selection, we ultimately chose the one that yielded the best results after trying several formats. However, since we did not cover all available prompt formats, the one we selected may not be the optimal one. 
Lastly, our benchmark includes GitHub data before 2023, which might already be present in the pretraining dataset of the models we studied, which poses a risk of data leakage. To mitigate this risk, we have implemented an automated benchmark construction pipeline that continuously updates the dataset and rewrited the docstrings for every task instance in \HumanEvo.

\begin{table}[t]
    \centering \small
    \setlength{\tabcolsep}{2pt}
    \caption{Model performance under different docstrings styles.} 
    \label{RQ3}
    \scalebox{0.9}{
    \begin{tabular}{l l rrrr}
    \toprule
        \multirow{3}{*}{Model} & \multirow{3}{*}{Docstring} & \multicolumn{2}{c}{Python} &  \multicolumn{2}{c}{Java}\\ 
        \cmidrule(lr){3-4} \cmidrule(lr){5-6}
       ~ & & EI & \HumanEvo &EI & \HumanEvo\\ \midrule
        \multirow{2}{*}{CodeLlama-7B} 
        &Detailed &29.0\%&22.0\%&8.5\%&6.0\%\\
        &Brief &28.5\%&22.0\%&7.0\%&4.5\%  \\\hline
        \multirow{2}{*}{CodeLlama-13B} 
        &Detailed &30.5\%&23.0\%&9.5\%&7.0\%\\
        &Brief &30.0\%&22.5\%&8.5\%&5.0\%  \\\hline
        \multirow{2}{*}{CodeLlama-34B} 
        &Detailed &31.5\%&24.0\%&14.5\%&10.0\%\\
        &Brief &31.0\%&24.5\%&11.5\%&7.0\%  \\\hline
        \multirow{2}{*}{DeepSeekCoder-6.7B} 
        &Detailed &30.0\%&23.0\%&13.0\%&9.0\%\\
        &Brief &30.0\%&23.0\%&10.5\%&6.5\%  \\\hline
        \multirow{2}{*}{DeepSeekCoder-33B} 
        &Detailed &31.0\%&25.0\%&15.5\%&11.5\%\\
        &Brief &31.0\%&24.0\%&13.5\%&9.0\%  \\\hline
        \multirow{2}{*}{GPT-3.5-Turbo} 
        &Detailed &32.5\%&25.5\%&18.0\%&13.0\%\\
        &Brief &31.5\%&24.5\%&14.5\%&11.0\%  \\\hline
        \multirow{2}{*}{GPT-4} 
        &Detailed &34.5\%&26.5\%&20.5\%&14.5\%\\
        &Brief &32.5\%&25.0\%&16.5\%&12.5\%  \\\bottomrule
     
    \end{tabular}
    }
\end{table}

\section{Related Work}
\subsection{Code Generation Benchmarks}
Previously, the vast majority of benchmarks were constructed to evaluate the performance of LLMs in generating standalone functions~\cite{multilingual, cassano2022scalable,jain2022jigsaw,ijcai2022p329, classeval, crosscodeeval, xcodeeval,chen2024rmcbench,zhang2024llm,zhong2024memorybank}, with the most representative ones being HumanEval ~\cite{humaneval} and MBPP ~\cite{MBPP}. HumanEval comprises 164 manually crafted programming problems covering language comprehension, reasoning, algorithms, and simple mathematics. MBPP consists of 974 short Python function problems primarily designed for novice programmers. 

Finding previous benchmarks fail to reflect complex programming scenarios, researchers propose CoderEval~\cite{codereval}, the first repository-level code generation benchmark, to evaluate the performance of existing LLMs in real-world development by prompting them to generate functions selected in actual complex software. CoderEval meticulously selects 230 Python and 230 Java programming tasks from popular real-world open-source projects, evaluating the functional correctness of generated code based on execution. EvoCodeBench~\cite{evocodebench} collects data from 25 real-world repositories. It also ensures that the distribution of programming tasks (with or without dependencies) matches the distribution of real repositories. RepoCoder~\cite{repocoder} proposed RepoEval, which consists of the latest and high-quality real-world repositories covering scenarios such as line completion, API invocation, and function body completion. Currently, most benchmarks for LLM code generation provide simple functions, mostly targeting general functions rather than specific domains, thus limiting practicality. However, BioCoder~\cite{biocoder}, a benchmark for bioinformatics code generation, has been introduced. BioCoder carefully selects 28 high-quality projects related to bioinformatics, then parses these repositories to generate ASTs and obtain relevant data. A custom code context is written for each filtered function, including necessary imports, cross-file dependencies, and some test cases.

However, all the benchmarks mentioned above have a common flaw: they all overlook the evolution of projects over time. These benchmarks take the latest version of the project as the context source, and such an evolution-ignored setting makes it difficult for them to accurately reflect the performance of LLMs in real-world development scenarios. Therefore, in this work, we propose \HumanEvo, the first repository-level evolution-aware code generation benchmark. Before obtaining the project context for the target function, we roll back the repository to the state before the target function was committed to ensure that the project context provided to LLMs does not suffer from issues such as future context leakage and missing useful context. Our objective is to offer LLMs a more realistic development scenario,  thus more accurately reflecting the actual performance of LLMs in real-world development scenarios.

\subsection{Repository-level Code Generation}
Many technical studies have been proposed to better tackle repository-level code generation tasks~\cite{repofusion,docprompting,repocoder,zan2023private,zan-etal-2022-language,bairi2023codeplan,repofuse,eghbali2024dehallucinator,zhang2024codeagent,phan2024repohyper,wang2024teaching,guo2024stop,chen2024identifying}.  
DocPrompting~\cite{docprompting} uses natural language (NL) intent to retrieve documents related to the target code and then sends NL intent and retrieved documents to LLMs. CoCoMIC~\cite{cocomic} develops a cross-file context finder, which is a static code analysis tool used to retrieve the most related cross-file context. MGD~\cite{guiding} leverages IDEs to assist in providing context. IDE is invoked to provide useful suggestions, when reaching the pre-defined trigger point. RLPG~\cite{repositorylevel} analyses dependencies potentially existing in the target function, then extracts project context that may be conducive to generating the target function. RepoCoder~\cite{repocoder} first slices the source code files in the repository to construct a retrieval library, then uses the generated problem as a query, calculates similarity to obtain project context, and feeds the project context and problem to the model to generate initial results. After obtaining the initial results, the previous generation results are used as queries for further querying to obtain more context. This process is repeated to continuously optimize the generation results. $A^3$-CodGen~\cite{codgen} first constructs a repository knowledge base covering knowledge of third-party libraries and extracts information on mobile and code context. During the code generation process, LLM first generates an initial version of code based on NL intent, then uses NL intent and the initial version of code to retrieve as much foundational knowledge as possible from the knowledge base for the LLM. Finally, local, global, and third-party library information is integrated into the prompt to allow the model to generate the final results.

All the mentioned works treat the latest version of the project as the context source, falling into the category of evolution-ignored scenarios. Although in RepoCoder, researchers begin to take notice of this issue by considering all code after the target function in the corresponding file as ``future context'', this still falls short of reality and introduces a lack of the context that should be present. Therefore, we propose a novel repository-level code generation benchmark, \HumanEvo, an evolution-aware pragmatic benchmark, to better simulate real development scenarios for LLMs.

\subsection{LLMs for Code Generation}

CodeLlama~\cite{codellama} is further trained based on Llama2~\cite{llama2}. It is capable of understanding and generating code in various programming languages, including but not limited to Python, Java, JavaScript, etc. CodeLlama not only generates complete code but also provides suggestions and optimization solutions for specific code snippets, greatly assisting developers. During the training of the base model, weights are initialized with an equivalent parameter amount of the Llama2 model, followed by training on a dataset of 500 billion tokens of code. 
StarCoder~\cite{starcoder} is derived from StarCoderBase, which is trained using 10 trillion tokens from The Stack, a vast collection of licensed GitHub repositories with inspection tools and selection exit procedures.
The DeepSeek-Coder~\cite{deepseek} takes into account the often intricate dependency relationships inherent in real-world projects. The model needs to fully consider these dependencies before completing programming tasks. Therefore, to equip DeepSeek-Coder to handle real-world complex projects, the training data is structured to support repository-level comprehension, enhancing the model's understanding of complex projects. This enables the model to utilize cross-file context effectively. 
MetaGPT~\cite{metagpt} constructs a multi-agent collaborative framework, employing Standardized Operating Procedures and involving multiple roles such as programmers and product managers to collaborate on project completion. The results generated by agents are not saved in conversational form but in structured output formats such as charts, files, etc. Communication between agents is also conducted through structured files. MetaGPT also implements an iterative development pipeline, optimizing results through multiple rounds of iteration by recording information generated during the development process by different roles. Similarly, ChatDev~\cite{chatdev} adopts a multi-agent framework, allocating and assigning different AI agents to various job functions. ChatDev facilitates communication among different role agents, forming a complete software development process where each agent performs its respective tasks while collaborating uniformly. In addition to these models, there are many other excellent models used in the field of code generation~\cite{wizardcoder,pangu,pangu2,codegen,codex,AlphaCode,incoder}.

\section{Conclusion}

In this paper, we highlight a previously overlooked flaw in evaluating existing repository-level code generation methods: the evolution of projects 
over time. To address it, we introduce \HumanEvo, an evolution-aware repository code generation benchmark, and conduct a comprehensive empirical study to explore LLMs' performance in real-world development scenarios. Experimental results demonstrate that the evolution-ignored situation leads to the inflated performance of LLMs. By incorporating temporal considerations into the benchmark, \HumanEvo provides a more comprehensive assessment of model performance in pragmatic development scenarios. We believe that our empirical results and \HumanEvo will offer valuable insights for the future development of LLMs-based repository-level code generation methods and benchmarks.


\section{Acknowledgements}
This work is supported by the National Natural Science Foundation of China (62032025), the Guangdong Basic and Applied Basic Research Foundation (2023A1515012292) and CCF-Huawei Populus Grove Fund CCFHuaweiSE202403.

\bibliographystyle{IEEEtran}
\bibliography{ref}

\end{document}